August 10, 2015

# A study on the shielding mechanisms of SOI pixel detector

Yunpeng Lu[a,b,*], Yi Liu[a,b,c], Zhigang Wu[a,b,d], Qun Ouyang[a,b], Yasuo Arai[e]

[a] State Key Laboratory of Particle Detection and Electronics, 100049 Beijing, CHINA
[b] Institute of High Energy Physics, CAS, 100049 Beijing, CHINA
[c] University of Chinese Academy of Sciences, 100049 Beijing, CHINA
[d] Department of Modern Physics, USTC, 230026 Hefei, CHINA
[e] Institute of Particle and Nuclear Studies, KEK, 305-0801 Tsukuba, JAPAN

ABSTRACT

In order to tackle the charge injection issue that had perplexed the counting type SOI pixel for years, two successive chips CPIXTEG3 and CPIXTEG3b were developed utilizing two shielding mechanisms, Nested-well and Double-SOI, in the LAPIS process. A TCAD simulation showed the shielding effectiveness influenced by the high sheet resistance of shielding layers. Test structures specially designed to measure the crosstalk associated to charge injection were implemented in CPIXTEG3/3b. Measurement results proved that using shielding layer is indispensable for counting type pixel and Double-SOI is superior to Nested-well in terms of shielding effectiveness and design flexibility.

PRESENTED AT

International Workshop on SOI Pixel Detector

(SOIPIX2015)

Tohoku University, Sendai, Japan, 3-6, June, 2015

___________________________

This work supported by National Natural Science Foundation of China (Grant No. 11375226), and in part by the CAS Center for Excellence in Particle Physics (CCEPP).
* Email: yplu@ihep.ac.cn

# 1. Introduction

The LAPIS process for SOI pixel detector allows fully depleted sensors and sub-micron CMOS circuits integrated in a single chip, which delivers high performance of particle detection[1]. However, the strong electrical coupling between pixel circuits and sensors remains a fundamental issue. The influence of sensors to the threshold of transistors, or known as the back-gate effect, was solved by the introduction of BPW. But the reverse influence of pixel circuits to sensors, or called charge injection in this article, was never solved properly. The impact of charge injection can be illustrated by a simple estimate as follow. Assuming a capacitance of $0.2fF/um^2$ between the device layer and the sensor, a voltage step of 1V on a square of $1um^2$ induces 1200 $e^-$ in the sensor, which is equivalent to a 4.3keV X-ray quantum.

The Fermilab group had aroused attention to the charge injection issue and proposed the Nested-well structure which became the first shielding mechanism for the LAPIS process[2][3]. The second shielding mechanism, Double-SOI, was developed by a collaborative effort of the KEK group and LAPIS[4]. The IHEP group started to work on counting type pixel in 2013 and paid a special attention to the charge injection issue based on the experience shared by the Fermilab and KEK group. Two successive chips CPIXTEG3 and CPIXTEG3b were developed by IHEP and KEK collaboratively, the former one based on Nested-well and the latter one based on Double-SOI.

# 2. Simulation

The shielding layers both in Nested-well and Double-SOI cannot be treated as ideal shielding conductor because a sub-micron thickness leads to high sheet resistance of 10kΩ/square for Nested-well and 30kΩ/square for Double-SOI. The sandwiched device layer, shielding layer and charge collection electrode in sensor comprise a RC network. In order to understand the crosstalk incurred by charge injection, a simplified 2D model was constructed for TCAD simulation, as showed in fig. 1. The Nested-well structure consists of a 20um-wide shielding layer grounded by the left contact, and a 23um-wide BPW as a charge collection electrode grounded by the right contact through a resistor of 800kΩ. The resistor here is to simulate the input impedance of a charge sensitive preamplifier which is commonly used in counting type pixel. There is a 1um-wide active area in the device layer used as a source of interruption.

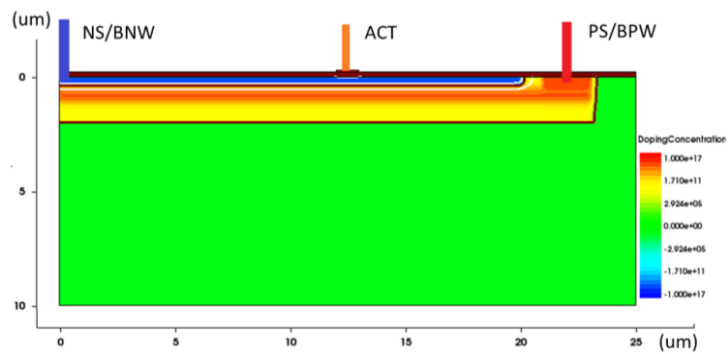

Fig. 1. The Nested-well structure used in TCAD simulation.

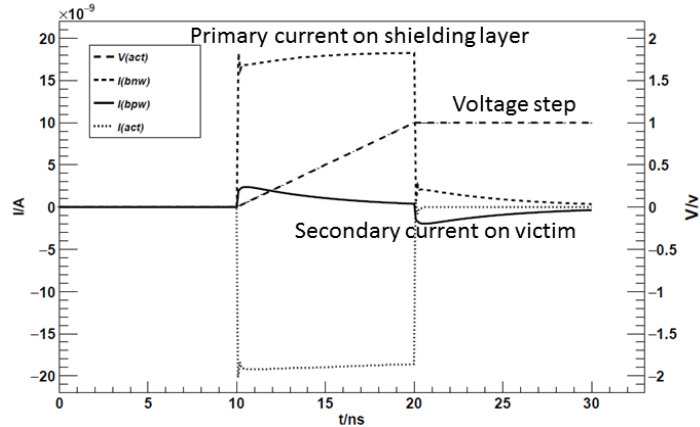

Fig. 2. The transient waveform in TCAD simulation.

Fig. 2 shows the voltage step used as a stimulus and corresponding transient current induced by it. The voltage step has amplitude of 1V and a rising edge of 10ns. The amplitude of induced current on the shielding layer is determined by the slope of dV/dt. Due to the sheet resistance of shielding layer, this current generates a voltage drop, which induces a secondary current on the charge collection electrode. The primary current on the shielding layer is proportional to dV/dt and decreases significantly when the voltage step attains constant, which leads to the disappearance of voltage drop on the shielding layer and induces the secondary current of inverse polarity on the charge collection electrode. The transient current on the active area, I(ACT), always equals to the sum of I(BNW) and I(BPW) as showed in fig.2. The shielding layer takes a large portion of injected charge. The charge collection electrode only takes the secondary current which can be decreased further by reducing the sheet resistance of shielding layer. It should be noted that the secondary current is a short bipolar pulse and charge sensitive preamplifiers are insensitive to it because its net charge is 0.

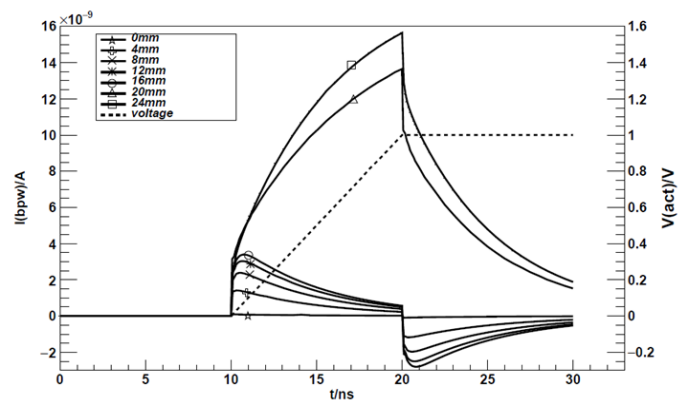

Fig. 3. The induced current on the charge collection electrode changes as the active area moves away from the BNW contact.

The amplitude of secondary current is influenced by the position of interruption source as showed in fig. 3. As the active area moves away from the BNW contact, the primary current sees more resistance and generates larger voltage drop, therefore the secondary current gets larger. The direct capacitive coupling dominates when the active area gets close enough to the edge of BPW. The induced current on charge collection electrode becomes unipolar and much larger in this case. The edge of BPW that cannot be shielded by BNW is an obvious weak point.
In the case of Double-SOI however, the shielding layer SOI2 is independent of charge collection electrode and enables full coverage of it.

## 3. Test structures

In practice the situation in a realistic counting type pixel is much more complicated than in the TCAD model, therefore it is necessary to test the shielding effectiveness in a real pixel chip. Any crosstalk incurred by the charge injection of circuit action would manifest itself on the analog node of signal processing. Hence inspecting the output of preamplifier, shaper and discriminator by using an oscilloscope is a powerful tool to discern and measure the crosstalk. In addition, the crosstalk due to charge injection is correlated to the circuit action and repeats as the circuit action does, which distinguishes between charge injection and random pickup.

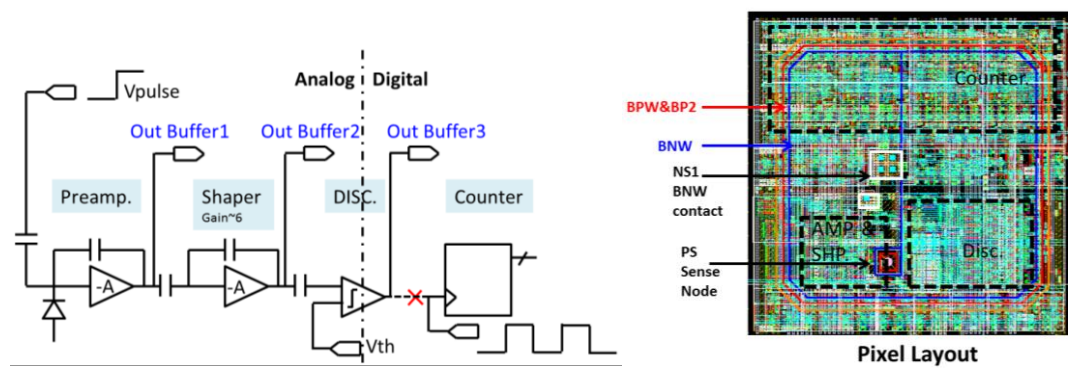

Fig. 4. Schematics (left) and Layout (right) of Nested-well test structures.

The test structure consists of a preamplifier, a shaper, a discriminator and a counter as showed in fig 4 (left). In order to measure the crosstalk arising from analog and digital separately, the output of discriminator was disconnected from the input of counter. The test input can be either a voltage step applied to the preamplifier or a digital clock applied to the counter. In the first case, any crosstalk from the analog circuit would distort the expected waveforms and in the second case any crosstalk from the digital circuit would generate unexpected output at the analog nodes. The pixel layout of CPIXTEG3 is showed in fig. 4 (right). Its nested-well is smaller than the pixel size of 64*64um. During the design of this chip, it's not perceived that the ring of BPW located outside of shielding layer would degrade the shielding effectiveness.

The test structure in CPIXTEG3b adopted similar pixel circuit and layout but its shielding layer, SOI2, had a full coverage on the entire pixel layout which is critical in improving the shielding effectiveness.

## 4. Test results

The CPIXTEG3 was manufactured in 2013 and tested intensively afterwards. Fig. 5 (left) is the output waveform of a Nested-well test structure with a voltage step applied to its preamplifier. The amplitude of voltage step is 600mV, equivalent to input charge of 14.4ke$^-$. All the waveforms of preamplifier, shaper and discriminator developed as expected except for a little distortion observed for the preamplifier which was influenced by the output stage of discriminator. The distortion can be eliminated by disabling the output stage of discriminator.
A comparative test was performed on a pixel cell without shielding layer in CPIXTEG3, in order to understand the crosstalk better. Fig. 5 (right) showed self-sustained oscillation on this pixel cell without any test input. The oscillation can be stopped by switching off the shaper. We speculated that the positive feedback path between shaper and sensor caused the oscillation. In fact, the self-oscillation could also be stopped by just decreasing the voltage gain of shaper. With a reduced factor of positive feedback, the output of shaper showed damped oscillation if a voltage step applied to the preamplifier. Obviously the shielding layer is absolutely necessary for the proper operation of conventional ASD signal processing.

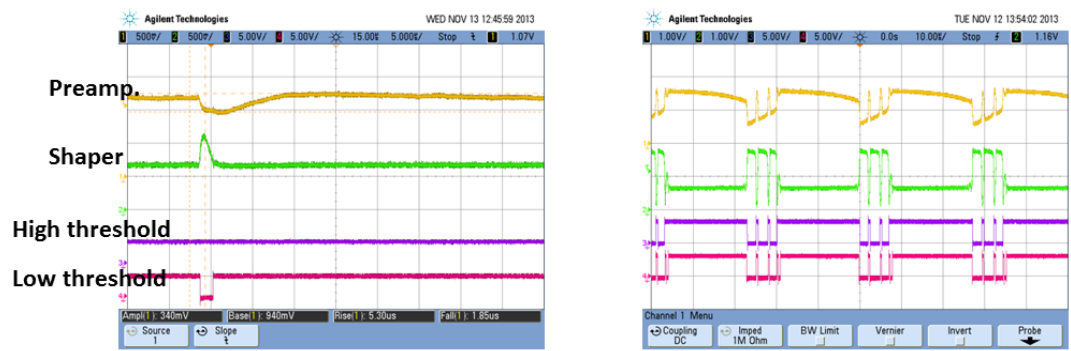

Fig. 5. Output waveform of Nested-well (left) and a single BPW without shielding layer (right).

However the counter in CPIXTEG3 generated large spurious signal as showed in Fig. 6 (left). The period of clock that drove the counter was 100us. The crosstalk correlated to the falling edge of clock was identical, because the output of counter didn't register any new state and the only logic change happened in the input stage of the lowest bit of counter. The crosstalk correlated to the rising edge of clock was obviously related to the output bit pattern. It reached maximum at the transition of all 1's to all 0's, with a spurious signal of 20ke$^-$ equivalent to input charge. We speculate that the crosstalk was caused by the outer ring of BPW that is not shielded at all. In contrast, Fig. 6 (right) is the analog output of CPIXTEG3b while the counter was driven by 100 kHz clock. It should be noted that the scale of amplitude for the preamplifier and shaper are 50mV/div and 100mV/div respectively. No discernable crosstalk correlated to any edge of clock was found in this case.

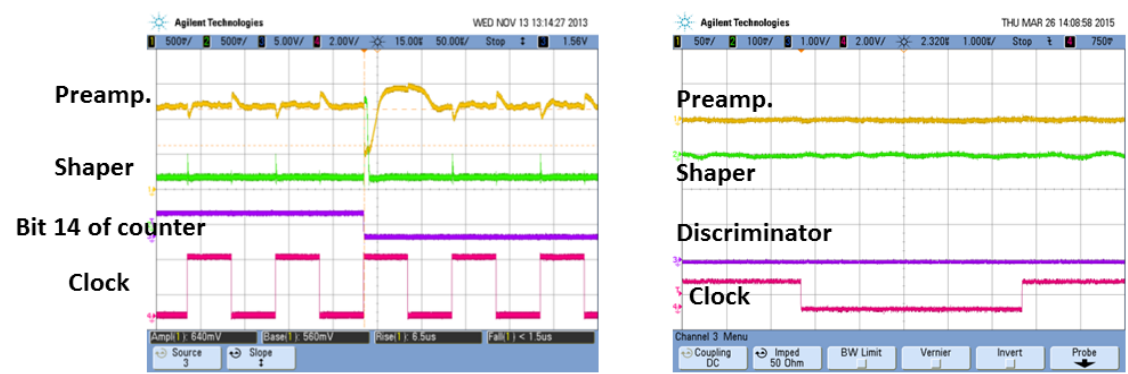

Fig. 6. Crosstalk induced by counter in CPIXTEG3 (left) and CPIXTEG3b (right).

## 5. Conclusion

Nested-well and Double-SOI are indispensable to counting type pixel, by which the crosstalk between pixel circuits and sensors can be shielded effectively.

Double-SOI is superior to Nested-well in two aspects. The first one is SOI2 can provide full coverage to shield the pixel circuit completely, while the edge of BPW in a nested-well structure has to expand out of shielding. Although a large enough Nested-well is likely to mitigate this problem, it is usually not practical in a pixel design that requires the highest density. The second one is the charge collection electrode of Double-SOI is independent on the shielding layer. It is possible to shrink the size of charge collection electrode to reduce the detector capacitance, which is critical to improve the signal noise ratio.

## Acknowledgements

We are grateful to the LAPIS Co. for being a supportive industrial partner to R&D of SOI pixel detector.